\documentclass{PoS}

\usepackage{amsmath}

\newcommand\iso[2]{\mbox{${}^{#2}${\rm #1}}}
\def\he#1{\iso{He}{#1}}

\def\li#1{\iso{Li}{#1}}
\def\b1#1{\iso{B}{1#1}}

\newcommand{\grav}{\widetilde{G}  }
\newcommand{\lsp}{\tilde\chi^0_1}
\newcommand{\mlsp}{m_{\tilde\chi^0_1}}
 
\newcommand{\mgrav}{m_{\widetilde{G}}}

\newcommand{\tanb}{\tan\beta}
\newcommand{\GeV}{\; \mathrm{GeV}}

\newcommand{\beq}{\begin{equation}}
\newcommand{\eeq}{\end{equation}}
\newcommand{\bea}{\begin{eqnarray}}
\newcommand{\eea}{\end{eqnarray}}

\def\b{\beta}

\title{GravitinoPack and late decays involving gravitinos}

\ShortTitle{GravitinoPack}

\author{Helmut~Eberl\\
   Institut f\"ur Hochenergiephysik der \"Osterreichischen Akademie der Wissenschaften,\\  A--1050 Vienna, Austria\\
        E-mail: \email{helmut.eberl@oeaw.ac.at}}

\author{\speaker{Vassilis~C.~Spanos}  \\  
        University of Athens, Faculty of Physics, Department of Nuclear {\rm \&} Particle Physics, \\ GR--15784 Athens, Greece\\
        E-mail: \email{vspanos@phys.uoa.gr}}

\abstract{In this talk, we present   the package {\tt {GravitinoPack}} that calculates  decays  of unstable supersymmetric 
particles, involving  gravitinos in the final or initial state.
If the gravitino is the dark matter particle and therefore stable, the package calculates the decays of the lightest neutralino, 
and the lighter stau or stop NLSP into the gravitino LSP and one or two Standard Model particles. 
On the other hand, assuming that the gravitino is unstable, {\tt {GravitinoPack}} calculates all its two-body and the three-body decay widths
to the neutralino LSP and Standard Model particles. Since all these decays, involving the gravitino, are of gravitational 
nature, the lifetime of the decaying particle can be of the order of seconds are more, hence called  ``late decays''.
The precise knowledge of all these partial decay widths enables the user to apply accurately the relevant cosmological constraints.}

\FullConference{18th International Conference From the Planck Scale to the Electroweak Scale \\
		25-29 May 2015\\
		Ioannina, Greece }

\begin{document}

\section{Introduction}
\label{intro}

The supersymmetric partner of the graviton, the gravitino,
belongs to the   spectrum of  models of particle physics, 
that incorporate the local version of  supersymmetry, the so-called supergravity.
Depending on the mass hierarchy of these models, 
in the case of $R$-parity conservation,  
the gravitino can be either the stable 
lightest supersymmetric particle  (LSP)  that  plays the role
of the dark matter (DM) particle, or it is heavier than the LSP and thus unstable. 

If the gravitino $\widetilde G$ is the LSP, other supersymmetric particles 
as neutralinos and sfermions (e.g. stops or staus) are unstable and can decay into a
gravitino and other Standard Model (SM) particles. 
These decays produce electromagnetic energy and hadrons which
affect the primordial Big-Bang
Nucleosynthesis (BBN) prediction~\cite{decays_old} for the  abundances of the light nuclei, 
like D, \he4, \he3 and \li7~\cite{decays_new, cefo_etc, cefos}.
On the other hand if the gravitino is not stable it can decay to the lightest neutralino ($\lsp$), 
that is the LSP in this case,  and other SM particles.

For both cases we have presented results, for the unstable~\cite{Gravitino_decays} and
the stable~\cite{GravitinoPack} gravitino, using the software {\tt GravitinoPack}~\cite{GravitinoPack}.
This package is a numerical tool which we developed with the help of the  packages
{\tt  FeynArts} ({\tt FA}), that was extended 
in order to deal with interactions with spin-3/2 particles with the particles of the Minimal Supersymmetric Standard Model (MSSM),
 and {\tt FormCalc} ({\tt FC})~\cite{FAref,FAFCref,LTref}, by using the Weyl-van-der-Waerden 
formalism~\cite{WvdW} as implemented into {\tt FC} from~\cite{Dittmaier:1998nn}. 
 {\tt GravitinoPack} contains {\sc Fortran77}  and  {\sc Mathematica} routines that calculate the decay widths for the
relevant  decay channels.

In~\cite{Gravitino_decays} we studied in detail all dominant two-body 
channels $\grav \to \widetilde{X}\, Y$, as
well as the three-body channels  $ \grav  \to  \lsp \, X \,Y $,   where $\widetilde{X}$ is a sparticle, $\lsp$ the lightest 
neutralino and $X$, $Y$ are SM particles. The two-body
decays dominate the total gravitino width, and in particular the channel $\grav \to  \lsp \, \gamma $,  which  is kinematically open 
in the whole region $\mgrav > \mlsp$. On the other hand, 
also many three-body decay channels can be open, $ \grav  \to  \lsp \, X \,Y $, even below thresholds of involved
two-body decays, $\mgrav < m_{\widetilde{X}}  + m_Y$.
For the gravitino DM (GDM) models (stable gravitino) 
the Next to the Lightest  Supersymmetric Particle (NLSP) can be the lightest neutralino ($\lsp$), 
or the lighter stau ($\tilde \tau_1$), or the lighter stop ($\tilde t_1$).
In this cases {\tt GravitinoPack} calculates all two-body decay channels  
e.~g. the dominant channel $\lsp \to \grav \, \gamma$, or $ \tilde \tau_1 \to  \grav \, \tau  $ or $ \tilde t_1 \to  \grav \, t $,
as well as all possible three-body decays.
 
As numerical applications of the package we  study a few representative 
benchmark points from supersymmetric models with different supersymmetry breaking patterns,
like the phenomenological MSSM (pMSSM)~\cite{pMSSM} and the 
Constrained MSSM (CMSSM)~\cite{cmssm}. For the pMSSM especially, we have selected 
points where the neutralino carries significant Higgsino components as in the  
Non-Universal Higgs Model (NUHM)~\cite{nuhm}. 

\section{The Decays in the {\tt \bf GravitinoPack}}
The main aspects were already presented in \cite{Gravitino_decays} 
and in the Appendix of~\cite{GravitinoPack}.  
There, one can find a detailed derivation of the 78 gravitino couplings with the
MSSM particles. Furthermore, in \cite{GravitinoPack} also the manual of {\tt GravitinoPack} is included.
The partial width of all decay channels presented in this section can be calculated with {\tt GravitinoPack}.

 \label{sect:gravdecays}
\subsection{Two-body decays of \boldmath $\widetilde G$}
\noindent
The gravitino decays into a fermion $F$, and a scalar $S$ or a vector particle $V$ are
\begin{eqnarray}
\widetilde G & \to & F  \,  S  \nonumber \\
& \to & f \tilde f^*_i,\, \bar f \tilde f_i , \,  \tilde \chi^0_j   \,  (h^0, H^0, A^0),\, \tilde \chi^+_k  \,  H^-,\, \tilde \chi^-_k  \,  H^+ ,  \nonumber \\
\widetilde G & \to & F  \,  V  \nonumber \\
& \to & \tilde g  \,  g,\, \tilde \chi^0_j  \,  (Z^0, \gamma),\, \tilde \chi^+_k   \,  W^- ,\, \tilde \chi^-_k   \,  W^+ ,
\label{two-body decays}
\end{eqnarray}
where $f$ denotes a SM fermion, $f  =  \nu_e, \,\nu_\mu,\, \nu_\tau,\,  e^-,\, \mu^-, \, \tau^-, \,  u,\, c,\, t,\, d,\, s,\, b$.
Its corresponding superpartners are the sfermions $\tilde f_i$, $i = 1,2$. 
The four neutralino states are 
 $\tilde \chi^0_j$, $j = 1,\ldots,4$ and the two charginos 
$\tilde \chi^\pm_k$, $k = 1, 2$. With $g$ we denote the gluon and with $\tilde g$ its superpartner, the gluino.
Furthermore, in the MSSM we have
three neutral Higgs bosons (two $CP$-even:  $h^0$ and $H^0$, and one  $CP$-odd: $A^0$) and two charged 
Higgs bosons: $H^\pm$. The vector bosons are the photon $\gamma$, the $Z$-boson $Z^0$ and the $W$-bosons $W^\pm$.

\subsection{Three-body decays of \boldmath $\widetilde G$}
\noindent
The gravitino decays into a neutralino and a pair of SM particles are
\begin{eqnarray}
\widetilde G & \to  & \tilde \chi_i^0 \, \bar f f\, ,  \nonumber \\
\widetilde G & \to  & \tilde \chi_i^0 \, V V\, ,\quad V V = Z^0 Z^0\,, Z^0 \gamma\,, W^+ W^- \, ,\nonumber \\
\widetilde G & \to  & \tilde \chi_i^0 \, V S\, ,\quad V S = (Z^0, \gamma)(h^0, H^0, A^0), W^+ H^-, W^- H^+\, ,\nonumber \\
\widetilde G & \to  & \tilde \chi_i^0 \, S S\, ,\quad S S = (h^0, H^0, A^0) (h^0, H^0, A^0), H^+ H^-\,.
\label{three-body decays}
\end{eqnarray}
These are 19 three-body  decay channels. 
 Note, that $\widetilde G \to \tilde \chi^0_i  \,  W^-  \,  H^+$ and its charge conjugated process
$\widetilde G \to \tilde \chi^0_i  \,  W^+  \,  H^-$ count as individual contributions, but
$\Gamma(\widetilde G \to \tilde \chi^0_i   \, W^-  \,  H^+) =  \Gamma(\widetilde G \to \tilde \chi^0_i  \,  W^+  \,  H^-)$.

\begin{figure}[h!]
\begin{center}
\resizebox{14cm}{!}{\includegraphics{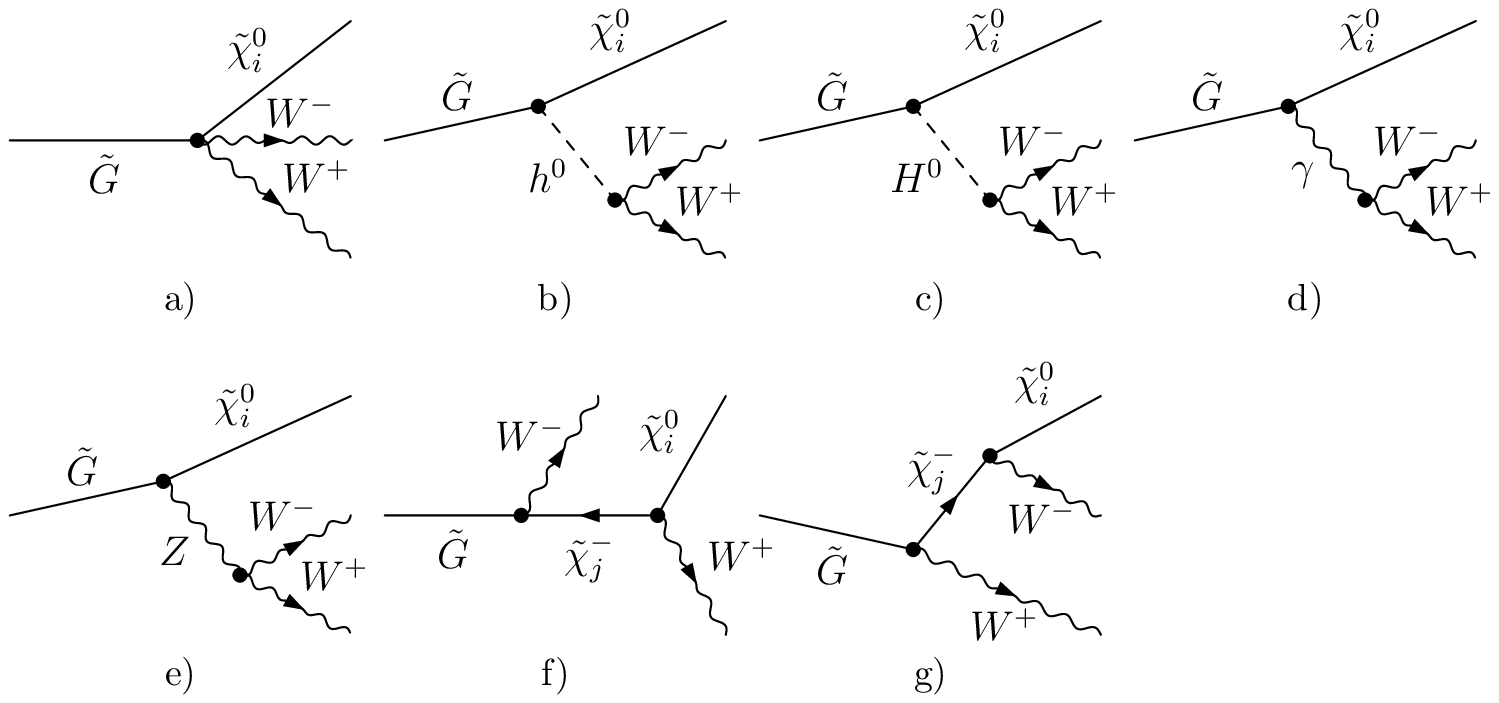}}
\end{center}
\caption{Feynman graphs for the decay ${\widetilde G} \to \tilde \chi^0_i W^- W^+$.
\label{fig:eynGr2NeuWmWp}}
\end{figure}

We  use as an example  the process  ${\widetilde G} \to \tilde \chi^0_i  \,  W^-  \,  W^+$ 
 to illustrate  the  nine individual amplitudes  contributing to this channel, as plotted  in  Figure~\ref{fig:eynGr2NeuWmWp}.

Now we turn to the stable gravitino scenario, discussing the decays of the lightest neutralino  $\widetilde \chi_1^0$
to gravitino and SM particles, followed by the discussion on the stau and stop NLSP decays. 

\subsection{ $\widetilde \chi_1^0$ decays into  \boldmath $\widetilde G$}
\label{sect:neudecays}
\noindent
Five two-body decays are possible:
\begin{eqnarray}
\widetilde \chi_1^0  & \to  & \widetilde G \,(Z^0\, ,\gamma) \, ,\nonumber \\
\widetilde \chi_1^0  & \to  & \widetilde G \,(h^0, H^0, A^0)  \, .
\label{neu2Gr two-body decays}
\end{eqnarray}
\noindent
The lightest neutralino can decay into the gravitino and a pair of SM particles as
\begin{eqnarray}
\widetilde \chi_1^0  & \to  & \widetilde G \, \bar f f \,, \nonumber \\
\widetilde \chi_1^0  & \to  & \widetilde G \, V V\, ,\quad V V = Z^0 Z^0\,, Z^0 \gamma\,, W^+ W^- \, ,\nonumber \\
\widetilde \chi_1^0  & \to  & \widetilde G \, V S\, ,\quad V S = (Z^0, \gamma)(h^0, H^0, A^0), W^+ H^-, W^- H^+\,, \nonumber \\
\widetilde \chi_1^0  & \to  & \widetilde G \, S S\, ,\quad S S = (h^0, H^0, A^0) (h^0, H^0, A^0), H^+ H^-\,.
\label{neu2Gr three-body decays}
\end{eqnarray}
They are the same as given in eq.~(\ref{three-body decays}) with neutralino and gravitino interchanged. 

\subsection{$\tilde \tau_1^-$ decays into  \boldmath $\widetilde G$}
\noindent
The only possible stau NLSP two-body decay is
\begin{eqnarray}
\tilde \tau_1^- & \to  & \widetilde G \,\tau^-  \, .
\label{stau2Gr two-body decays}
\end{eqnarray}
\noindent
The  corresponding three-body decays are  
\begin{eqnarray}
\tilde \tau_1^-  & \to  & \widetilde G \, Z^0 \, \tau^- \nonumber \, ,\\
\tilde \tau_1^-  & \to  & \widetilde G \, W^-\, \nu_\tau \nonumber \,,\\
\tilde \tau_1^-  & \to  & \widetilde G \, (h^0, H^0, A^0)\, \tau^- \nonumber \,,\\
\tilde \tau_1^-  & \to  & \widetilde G \, H^-\, \nu_\tau \,.
\label{stau2Gr three-body decays}
\end{eqnarray}

\subsection{$\tilde t_1$  decays into \boldmath $\widetilde G$}
\noindent
There is only one stop NLSP two-body decay possible,
\begin{equation}
\tilde t_1  \to   \widetilde G \,t  \, .
\label{st2Gr2body decays}
\end{equation}
\noindent
The possible three-body decays of $\tilde t_1$ are 
\begin{eqnarray}
\tilde t_1 & \to  & \widetilde G \, Z^0 \, t\nonumber \, ,\\
\tilde t_1  & \to  & \widetilde G \, W^+\, b \nonumber \,,\\
\tilde t_1  & \to  & \widetilde G \, (h^0, H^0, A^0)\, t \nonumber \,,\\
\tilde t_1  & \to  & \widetilde G \, H^+\, b \,.
\label{stop2Gr three-body decays}
\end{eqnarray}

\section{Numerical  results}

In our numerical analysis we choose representative points from various supersymmetric models, assuming 
different  mechanism for  the supersymmetry breaking. In  particular, we study points based 
on the pMSSM~\cite{pMSSM}  and in   the 
CMSSM~\cite{cmssm}. 
The benchmark points we study in this section are compatible with the cosmological~\cite{planck,xenon100} 
and LHC constraints (Higgs mass $\simeq 125 \GeV$,
 LHCb bounds for rare decays etc.)~\cite{spheno,FH}.

\begin{figure}[t!]
\mbox{\hspace{2mm}
\includegraphics[width=7.7cm]{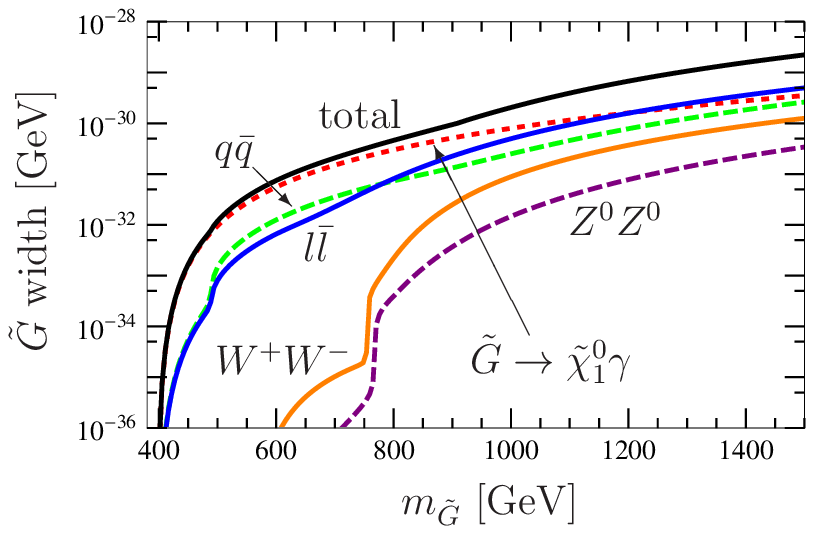}
\hspace{3mm}
\hfill
\includegraphics[width=7.6cm]{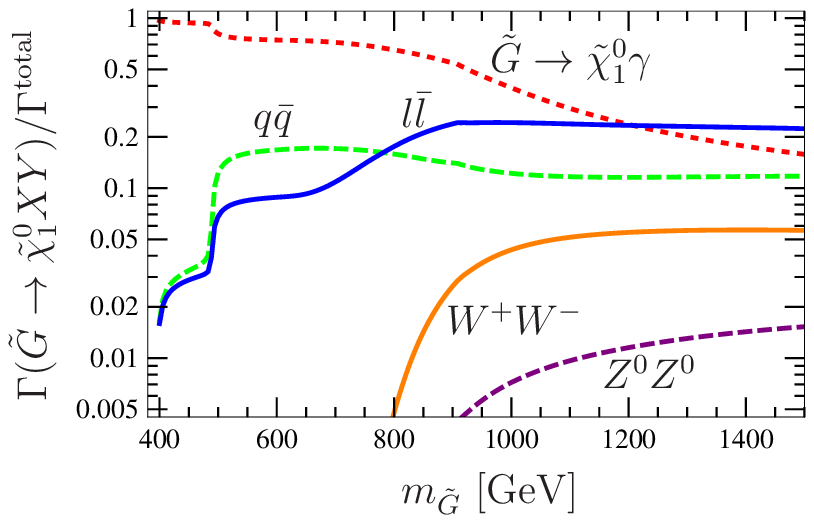}
}
\vspace*{-0.5cm}
\caption[]{The three-body decay widths of the gravitino decaying into $\lsp$ and $q \bar q$, $l \bar l$, $W$-pairs, and $Z$-pairs at the $A^0$-funnel point;
``total'' denotes $\Gamma^{\rm total}$ which is the full two-body width plus the sum of the non-resonant part of three-body decay widths with $\lsp$;
$q \bar q$ stands for the sum over all six quark flavours and $l  \bar l$ for the sum over the three charged lepton and three neutrino flavours.
The red dotted lines denote the two-body decay $\widetilde G \to \lsp \gamma$.  In the right figure we display the corresponding branching ratios for 
the decay channels plotted in the left figure.}
\label{fig:NeuDec_funnel}
\end{figure}

Our first numerical example is in the context of neutralino DM models with unstable gravitino, using 
 a pMSSM point with the SUSY parameters chosen as
 $\tan\beta=20$, $\mu=700$ GeV, $M_A=770 \GeV$, $(M_1,M_2,M_3) = (400,800,2400) \GeV$,
 $A_t=A_b=-2050$ GeV, $A_\tau=-1000$ GeV, 
 $m_{\widetilde{q}_{L}}=m_{\widetilde{u}_{R}}=m_{\widetilde{d}_{R}}=m_{\widetilde{b}_{R}}=1500$ GeV,
 $ m_{\widetilde{\ell}_{L}} = 600   \GeV$, $m_{\widetilde{e}_{R}}=2500 \GeV$, 
 $m_{\widetilde{Q}_{3L}}=m_{\widetilde{L}_{3L}}=800 \GeV$ and $m_{\widetilde{\tau}_{R}}=2000 \GeV$.
 In this point the neutralino relic density lies cosmologically along the so-called $A^0$-funnel region. 
Thus, we show in Figure~\ref{fig:NeuDec_funnel}
the three-body decay widths of the gravitino decaying into $\lsp$ together with $q \bar q$, $l \bar l$, and $W, \,Z$-boson pairs there.
We also show the two-body decay channel $\widetilde G \to \lsp \gamma$ as a reference, because it is the dominant one for small $m_{\widetilde G}$.
In the left figure we also show $\Gamma^{\rm total}$ which is the full two-body width plus the sum of the non-resonant part of 
three-body decay widths with $\lsp$,
denoted by ``total''. In the right figure we show the relative quantities in terms of $\Gamma^{\rm total}$;
$q \bar q$ stands for the sum over all six quark flavours, $\sum_{i = 1,6} \Gamma^{\rm reso~+~non-reso }(\widetilde G \to \lsp q_i \bar q_i)$ and $l \bar l$,
$q_i = u, d, c, s, t, b$, and $l  \bar l$ the sum of the three charged lepton and three neutrino flavours, $\sum_{j = 1,3} \left(\Gamma^{\rm non-reso~+~reso}(\widetilde G \to \lsp l^+_j  l^-_j)\right.$ + $\left.\Gamma^{\rm non-reso~+~reso}(\widetilde G \to \lsp \nu_{l_j} \bar \nu_{l_j})\right) $, $l_j = e, \nu, \tau$. 
The decay width summing up the decays into all fermion pair can reach 38\%, into the $W$-boson pair 5.6\% and 
into the $Z$-boson pair 1.5\% in the shown range.
The analogous plots for the coannhilation point look similar but the decays into $W$- and $Z$-boson pairs are suppressed by about two orders of magnitude
due to the pure bino state of the LSP.

\begin{figure}[t!]
\mbox{\hspace{2mm}
\includegraphics[width=7.7cm]{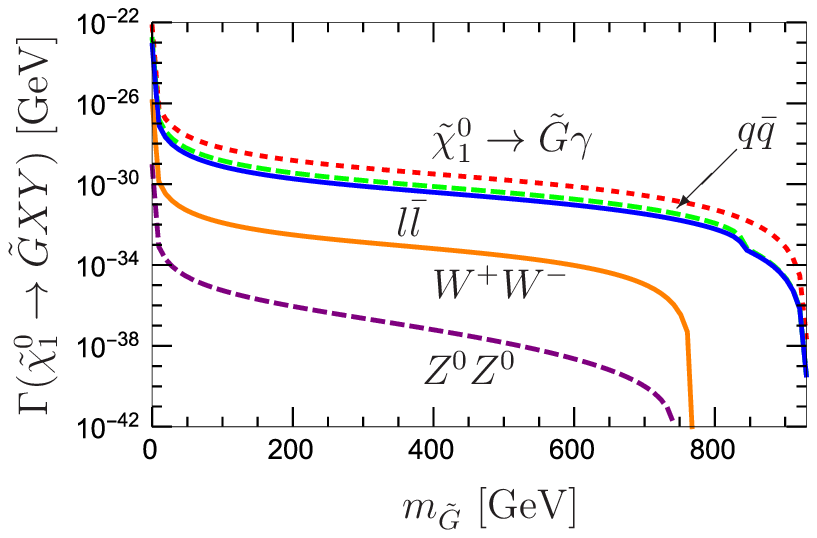}
\hspace{3mm}
\hfill
\includegraphics[width=7.6cm]{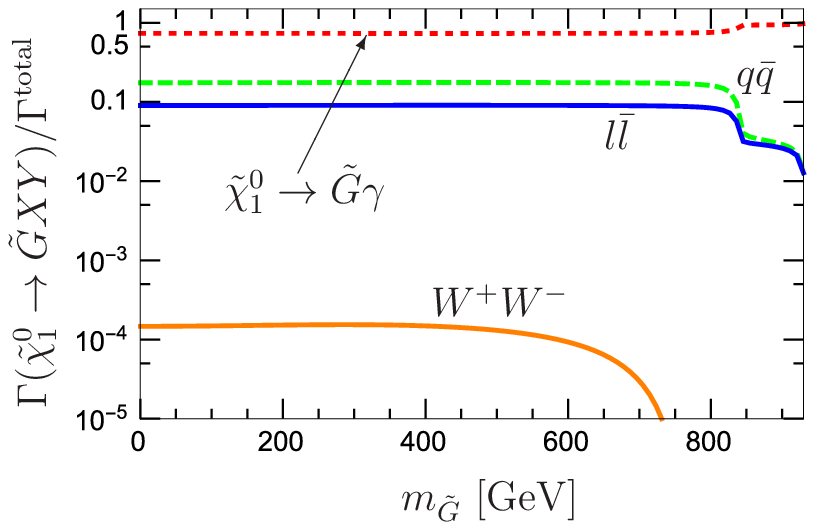}
}
\vspace*{-0.5cm}
\caption[]{The three-body decay widths of the neutralino NLSP decaying into the gravitino $\widetilde G$   and other particles, in the 
GDM/CMSSM scenario.  
The dominant channels  $q \bar q$, $l \bar l$, $W^+W^-$, and $ZZ$ are marked in the figure;
$q \bar q$ stands for the sum over all six quark flavours and $l  \bar l$ for the sum over the three charged lepton and three neutrino flavours.
The red dotted lines denote the two-body decay $ \lsp \to \widetilde G  \gamma$. 
 In the right figure we display the corresponding branching ratios for  the decay channels plotted in the left figure.}
\label{fig:neu_dec}
\end{figure}

The second numerical example we discuss is based on  a model with stable gravitino (GDM) in the CMSSM. 
We use $m_0=1600$, $M_{1/2}=5000$, $A_0=-4000$ GeV, and
$\tanb=10$. The mass of $\tilde {\chi}^0_1$ is 2282 GeV.
 It is worth mentioning that the gravitino DM relic density and the NLSP relic density 
 are related by
  \beq
 \frac{\Omega_{\mathrm {NLSP}}}{\Omega_{\widetilde G}}= \frac{m_{\mathrm {NLSP}} }{m_{\widetilde G}}  > 1\, .
 \eeq
The cosmological bound for the gravitino relic density can be understood as upper bound $\Omega_{\widetilde G} h^2 \lesssim 0.119$.
Therefore, one can have in addition gravitino production during reheating after inflation, if the reheating temperature is relatively large, 
of the order of $\sim 10^{10} \GeV$.

\begin{figure}[h!]
\mbox{\hspace{2mm}
\includegraphics[width=7.7cm]{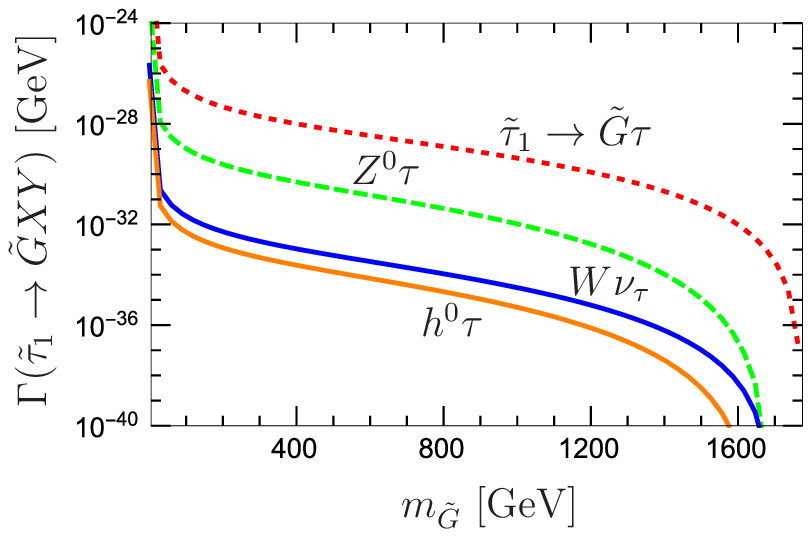}
\hspace{3mm}
\hfill
\includegraphics[width=7.6cm]{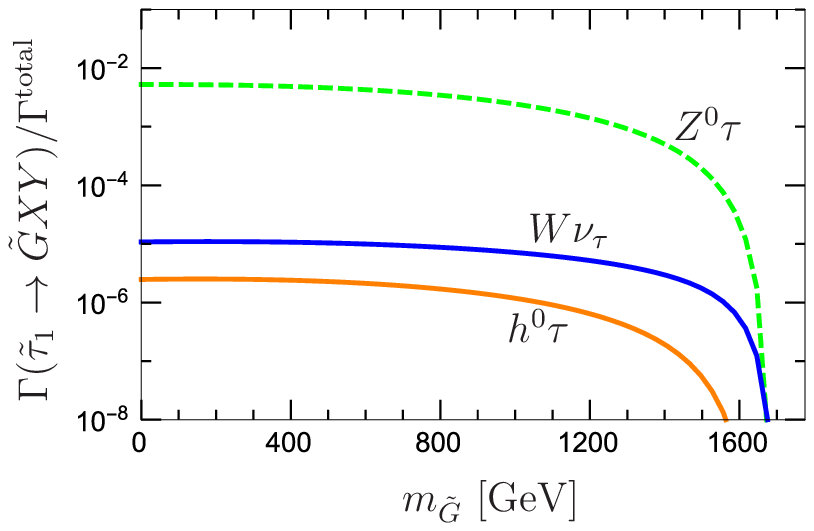}
}
\vspace*{-0.5cm}
\caption[]{The three-body decay widths of the stau  NLSP
 decaying into the gravitino $\widetilde G$  and other SM particles, in the GDM/CMSSM scenario. 
 We present the dominant two-body channel $\widetilde G  \tau $ and the three-body channels 
 $ \widetilde G Z \tau  $, $ \widetilde G W^- \nu_\tau  $ and $ \widetilde G h  \tau  $.
 In the right figure we display the corresponding branching ratios for 
the decay channels plotted in the left figure except $\tilde \tau_1\to  \widetilde G \tau$.}
\label{fig:stau_dec}
\end{figure}

\begin{figure}[t!]
\mbox{\hspace{2mm}
\includegraphics[width=7.7cm]{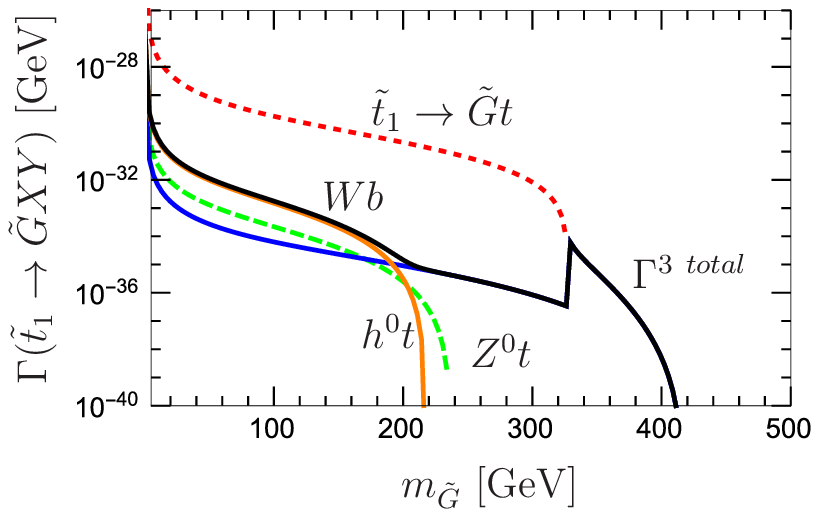}
\hspace{3mm}
\hfill
\includegraphics[width=7.6cm]{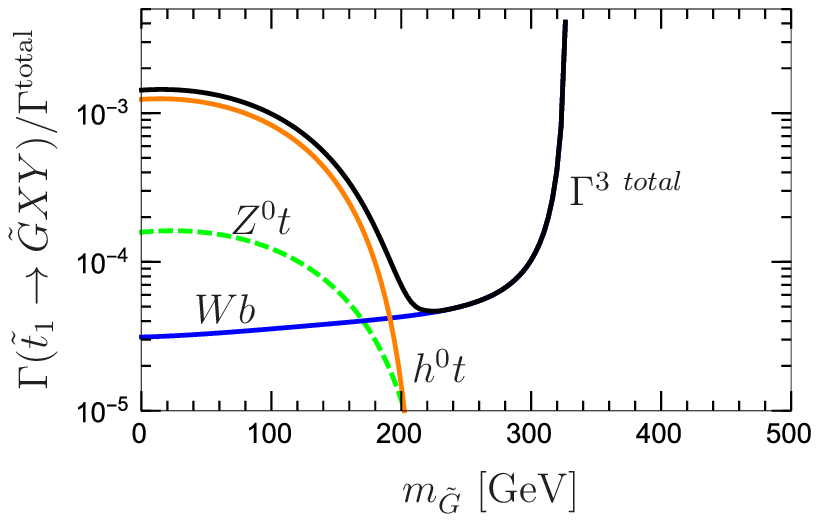}
}
\vspace*{-0.5cm}
\caption[]{The three-body decay widths of the stop  NLSP
 decaying into the gravitino $\widetilde G$  and other SM particles, in the GDM/CMSSM scenario.  
 We present the dominant two-body channel $\widetilde G  \tau $ and the three-body channels 
 $ \widetilde G Z t  $, $ \widetilde G W^- b  $ and $ \widetilde G h^0  b  $.
 In the right figure we display the corresponding branching ratios for 
the decay channels plotted in the left figure except $\tilde t_1\to  \widetilde G t$.}
\label{fig:stop_dec}
\end{figure}

In Figure~\ref{fig:neu_dec} we present the corresponding decay widths (left figure) and the branching ratios (right figure) for the 
neutralino decays into $\widetilde G$   and other particles.
The dominant channels  $q \bar q$, $l \bar l$, $W^+W^-$, and $ZZ$ are marked in the figure;
$q \bar q$ stands for the sum over all six quark flavours and $l  \bar l$ for the sum over the three charged lepton and three neutrino flavours.
The red dotted lines denote the two-body decay $ \lsp \to \widetilde G  \gamma$, that dominates the neutralino decay amplitude,
as can be seen in the left panel  that illustrates the branching ratios.
On the other hand, the  three-body decay channels  $q \bar q$ and $l \bar l$ are of the order of 10\%, while the
$W^+W^-$, and $ZZ$ channels are much smaller. For this particular CMSSM point the other decay channels are even smaller. 
This happens because the neutralino is predominantly a bino at this particular point of the parameter space. 

In Figure~\ref{fig:stau_dec} we present the  decays widths (left figure) and the branching ratios (right figure) for the 
stau NLSP decays into the gravitino and other particles, in the GDM scenario. 
The CMSSM parameters are $m_0=1000$, $M_{1/2}=4200$, $A_0=-2500$ GeV, and
$\tan\beta=10$. The mass of $\tilde\tau_1$ is 1795 GeV. The dominant decay channel is the 
two-body decay $\tilde\tau_1 \to \widetilde G \tau $. In addition we plot the three-body channels 
$\tilde\tau_1 \to  \widetilde G Z^0 \tau  $, $\tilde\tau_1 \to \widetilde G W^- \nu_\tau$, and $\tilde\tau_1 \to  \widetilde G h^0  \tau$. 
The widths of the channels involving heavier Higgs bosons in the final state, are much smaller or even zero.

Similarly, in Figure~\ref{fig:stop_dec} we present the  decays widths (left figure) and the branching ratios (right figure) for the 
stop NLSP decays into the gravitino $\widetilde G$ and other particles. 
The CMSSM parameters are
$m_0=3000$, $M_{1/2}=1090$, $A_0=-7500$ GeV and
$\tan\beta=30$. The  mass of the NLSP $\tilde t_1$ is 501~GeV.
In addition to $\tilde t_1 \to \widetilde G t $ we present 
also the three-body decay channels into
$\widetilde G Z^0 t$, $\widetilde G W^- b$ and $\widetilde G h^0  t$. 
Again, the channels involving the heavier Higgs bosons are negligible.
The dominant decay channel is $\tilde t_1 \to \widetilde G t $, up to the
kinematical threshold $m_{\widetilde G} = m_{\tilde t_1} - m_t \sim 328$~GeV.  
As can been seen in both plots in Figure~\ref{fig:stop_dec}, beyond this point the two-body channel is closed
and it dominates the three-body channel $\tilde t_1 \to \widetilde G W^- b$. 
This is clearer visible in the right plot, where for $m_{\widetilde G} > m_{\tilde t_1} - m_t$ the $ \widetilde G W^- b$
decay channel grows after this point and eventually reach the maximum value one outside of the displayed region. 
Recently, a paper studying in particular, this three-body channel appeared~\cite{arXiv:1510.01447}.

In  summary, we have discussed   representative cases both in the pMSSM and  CMSSM.
Based on these examples one can see that the full knowledge of the two- and three-body
decay channels of the NLSP is essential for the precise calculation of the decay width and the various 
branching ratios. This is actually the big advantage of using {\tt GravitinoPack}, since it gives all computed results both in 
{\sc Fortran77} and within the {\sc Mathematica} environment. It also supports SLHA input format.

\section{Summary}

We have studied supersymmetric models, where the gravitino is either unstable or stable. If it is
unstable it can decay to the neutralino LSP and SM particles.  
If it is the stable LSP it can play the role of the DM particle and the NLSP
can be the lightest neutralino $\lsp$ or a sfermion, as stau $\tilde \tau_1$ or stop $\tilde t_1$. 

In this talk we have discussed all these cases  using the recently presented  public available computer tool {\tt GravitinoPack}.
This numerical package based on an auto-generated {\sc Fortran77} code, calculates the branching ratios and decay widths for the 
NSLP decays, if the gravitino is stable and the DM particle. In this case we have  calculated 
decays of the NLSP neutralino, stau, and stop
to the gravitino LSP and one or two SM particles.
We have also discussed the complementary case,
where the gravitino is unstable and can decay into the neutralino LSP and SM particles.
The products of these decays carry electromagnetic energy and can build hadrons
that influence the predictions of BBN, since the gravitational nature of these decays place them in this time scale.  
Therefore, the detailed knowledge of the relevant branching ratios and decay widths is important to study and 
constrain various supergravity models.

\section*{Acknowledgements}

This work is supported by the Austrian Science Fund (FWF) P~26338-N27
and by the European Commission through the ``HiggsTool''  Initial Training Network PITN-GA-2012-316704.

\end{document}